\documentclass[aps,prc,twocolumn,showpacs,floatfix]{revtex4-1}

\usepackage{graphicx}
\usepackage{dcolumn}
\usepackage{bm}
\usepackage{ulem}
\usepackage{color}

\newcommand{\al}{$\alpha$}
\newcommand{\g}{$\gamma$}

\newcommand{\rag}{($\alpha$,$\gamma$)}
\newcommand{\ran}{($\alpha$,n)}
\newcommand{\rannull}{($\alpha$,n$_0$)}
\newcommand{\rani}{($\alpha$,n$_1$)}

\newcommand{\rap}{($\alpha$,$p$)}
\newcommand{\rapnull}{($\alpha$,$p_0$)}
\newcommand{\rpa}{($p$,$\alpha$)}
\newcommand{\rpanull}{($p$,$\alpha_0$)}
\newcommand{\rpai}{($p$,$\alpha_1$)}
\newcommand{\rpaii}{($p$,$\alpha_2$)}

\newcommand{\rng}{(n,$\gamma$)}
\newcommand{\rga}{($\gamma$,$\alpha$)}

\newcommand{\rna}{(n,$\alpha$)}
\newcommand{\rnanull}{(n,$\alpha_0$)}
\newcommand{\rnai}{(n,$\alpha_1$)}

\newcommand{\ciii}{$^{13}$C}
\newcommand{\ovi}{$^{16}$O}
\newcommand{\ovii}{$^{17}$O}
\newcommand{\oviii}{$^{18}$O}
\newcommand{\fvii}{$^{17}$F}
\newcommand{\nenull}{$^{20}$Ne}
\newcommand{\nei}{$^{21}$Ne}
\newcommand{\neii}{$^{22}$Ne}
\newcommand{\mgv}{$^{25}$Mg}

\newcommand{\Nsv}{$N_A$$\left< \sigma v \right>$}
\newcommand{\spro}{$s$-process}
\newcommand{\rpro}{$r$-process}

\newcommand{\gpro}{$\gamma$-process}

\newcommand{\sfact}{S-factor}

\begin{document}

\title{
  Uncertainty of the astrophysical $^{17,18}$O($\alpha$,n)$^{20,21}$Ne
  reaction rates and the applicability of the statistical model for nuclei
  with $A \lesssim 20$
}

\author{Peter Mohr}
\email[Email: ]{WidmaierMohr@t-online.de; mohr@atomki.mta.hu}
\affiliation{
Diakonie-Klinikum, D-74523 Schw\"abisch Hall, Germany}
\affiliation{
Institute for Nuclear Research (Atomki), H-4001 Debrecen, Hungary}

\date{\today}

\begin{abstract}
\begin{description}
\item[Background] The ($\alpha$,n) and ($\alpha$,$\gamma$) reactions on
  $^{17,18}$O have significant impact on the neutron balance in the
  astrophysical $s$-process. In this scenario stellar reaction rates are
  required for  relatively low temperatures below $T_9 \lesssim 1$.
\item[Purpose] The uncertainties of the $^{17,18}$O($\alpha$,n)$^{20,21}$Ne
  reactions are investigated. Statistical model calculations are performed to
  study the applicability of this model for relatively light nuclei in
  extension to a recent review for the $20 \le A \le 50$ mass range.
\item[Method] The available experimental data for the
  $^{17,18}$O($\alpha$,n)$^{20,21}$Ne reactions are compared to statistical
  model calculations. Additionally, the reverse $^{20}$Ne(n,$\alpha$)$^{17}$O
  reaction is investigated, and similar studies for the $^{17}$F mirror
  nucleus are provided.
\item[Results] It is found that on average the available experimental data for
  $^{17}$O and $^{18}$O are well described within the statistical
  model, resulting in reliable reaction rates above $T_9 \gtrsim 1.5$ from
  these calculations. However, significant experimental uncertainties are
  identified for the $^{17}$O($\alpha$,n$_0$)$^{20}$Ne(g.s.) channel.
\item[Conclusions] The statistical model is able to predict astrophysical
  reaction rates for temperatures above 1 GK with uncertainties of less than a
  factor of two for the nuclei under study. An experimental discrepancy for
  the $^{17}$O($\alpha$,n)$^{20}$Ne reaction needs to be resolved.
\end{description}
\end{abstract}

\maketitle

\section{Introduction}
\label{sec:intro}
\al -induced reactions play an important role in various astrophysical
scaenarios. In the astrophysical \spro , the \ciii \ran \ovi\ and \neii \ran
\mgv\ reactions are the neutron production reactions \cite{Kaepp11}, and the
\ovii \ran \nenull , \ovii \rag \nei , and \oviii \ran \nei\ reactions affect
the neutron balance via the potential neutron poison \ovi . Depending on
the rates of these reactions, a neutron may be first absorbed by the highly
abundant \ovi\ nucleus in the \ovi \rng \ovii\ reaction, but later the neutron
can be recycled in the \ovii \ran \nenull\ reaction \cite{Bar92,Mohr16b}.

In most cases the statistical model (StM) is applied for the calculation of
\al -induced reaction rates. The StM is well founded for heavy target nuclei,
e.g.\ for \ran\ reactions under certain \rpro\ conditions
\cite{Per16,Mohr16,Bli17}, and for inverse \rga\ reactions in the
\gpro\ \cite{Rau16,Sim17}. Contrary to the situations in the \rpro\ and \gpro
, it is not clear whether the level density is sufficiently high for a
reliable calculation of \al -induced reactions for the light target nuclei in
the \spro\ with masses $A \lesssim 20$. Interestingly, it was found that the
reaction cross sections in the $20 \le A \le 50$ mass range follow a generic
behavior and can be quite well described within the StM
\cite{Mohr15} using the simple 4-parameter potential by McFadden and Satchler
\cite{McF66}. This holds in particular for slightly higher energies and/or
nuclei at the upper end of the $20 \le A \le 50$ mass range. For low energies
and lighter target nuclei, the cross sections are dominated by individual
resonances, and thus the StM is only able to reproduce the
average trend of the experimental cross sections. Contrary to this excellent
performance for light target nuclei, the simple McFadden/Satchler potential
tends to overpredict \al -induced cross sections for heavy target nuclei in
the $A \approx 100$ mass range and above. Much effort has been spent in the
recent years to provide improved global \al -nucleus potentials for heavy
nuclei, and significant improvements have been achieved (e.g.,
\cite{Dem02,Mohr13,Avr14,Su15,Sch16}).

Primarily, this study was motivated as an extension of the previous review in
the $20 \le A \le 50$ mass range \cite{Mohr15} with the aim to provide a
prediction for the upcoming \fvii \rap \nenull\ data which have been measured
recently using the MUSIC chamber at Argonne National Lab
\cite{Avi18,Avi_priv}. A further experiment for \fvii \rap \nenull\ has been
done at Florida State University using the ANASEN active detector
\cite{Blackmon17}. In the course of this study of \fvii +\al , the mirror
\ovii \ran \nenull\ reaction was also analyzed, and unexpected inconsistencies
between different experimental data sets were identified. These are based on
the \ovii \ran \nenull\ data in
\cite{Han67,Bair73,McD76,Denker94,Best13_o17,Avi17} and the reverse \nenull
\rna \ovii\ reaction \cite{John51,Bell59,Khr12a,Khr12b}. As a consequence, the
present work now focuses on the resulting uncertainties of the \ovii \ran
\nenull\ reaction rate. In most cases
\cite{Han67,Bair73,Denker94,Best13_o17,Best13_o18} the same experimental
techniques were also applied to the \oviii \ran \nei\ reaction. This allows a
careful comparison of the experimental results for two nuclei, and in addition
a step-by-step extension of the systematics in the $20 \le A \le 50$ mass
range \cite{Mohr15} towards lighter nuclei is possible. Detailed calculations
of the \fvii \rap \nenull\ and its reverse \nenull \rpa \fvii\ reaction
\cite{Gru77} will be provided in a separate paper \cite{Avi18}. Most
experimental data in this work have been taken from the EXFOR database
\cite{EXFOR}; other sources are given explicitly.

The paper is organized as follows. Sec.~\ref{sec:o18} gives a review of the
existing experimental data for the \oviii \ran \nei\ reaction, and the
experimental data are compared to StM
calculations. Sec.~\ref{sec:o17} provides a similar review for the \ovii \ran
\nenull\ reaction which is extended by data for the reverse \nenull \rna
\ovii\ reaction. The \fvii \rap\ \nenull\ and \nenull \rpa \fvii\ reactions
are briefly mentioned in Sec.~\ref{sec:f17}. Astrophysical reaction rates are
calculated, and their uncertainties are discussed in
Sec.~\ref{sec:astro}. Finally, as the \al -nucleus potential is the key
ingredient for the calculation of \ran\ cross sections, the results from
different \al -nucleus potentials are presented in
Sec.~\ref{sec:alpha}. Conclusions are drawn in Sec.~\ref{sec:conc}.

The StM calculations in the present work were made using the code TALYS
\cite{TALYS} (version 1.8) in combination with the \al -nucleus potential by
McFadden/Satchler. This choice is based on the excellent performance of the
McFadden/Satchler potential in the $20 \le A \le 50$ mass range \cite{Mohr15}
and on the finding that most of the recent global potentials
\cite{Dem02,Avr14,Mohr13} provide relatively similar cross sections for
lighter nuclei \cite{Mohr17}. Other ingredients for the StM calculations like
the nucleon optical model potential, the \g -ray strength function, and the
level density have very minor influence on the calculated \ran\ or \rap\ cross
sections within the StM, in particular as long as either the \ran\ or the
\rap\ channel has a dominating contribution to the total \al -induced reaction
cross section; this is often the case in the $20 \le A \le 50$ mass range
\cite{Mohr15}. Although the role of the chosen level density parametrization
in the StM calculations is minor, in reality at the lowest energies under
study the cross sections are governed by the properties of few levels which
appear as low-energy resonances in the \ran\ cross section.

\section{\oviii ($\alpha$,${\rm{n}}$)\nei }
\label{sec:o18}
\begin{figure}[htb]
\includegraphics[width=0.99\columnwidth,bbllx=3,bblly=0,bburx=246,bbury=313,clip=]{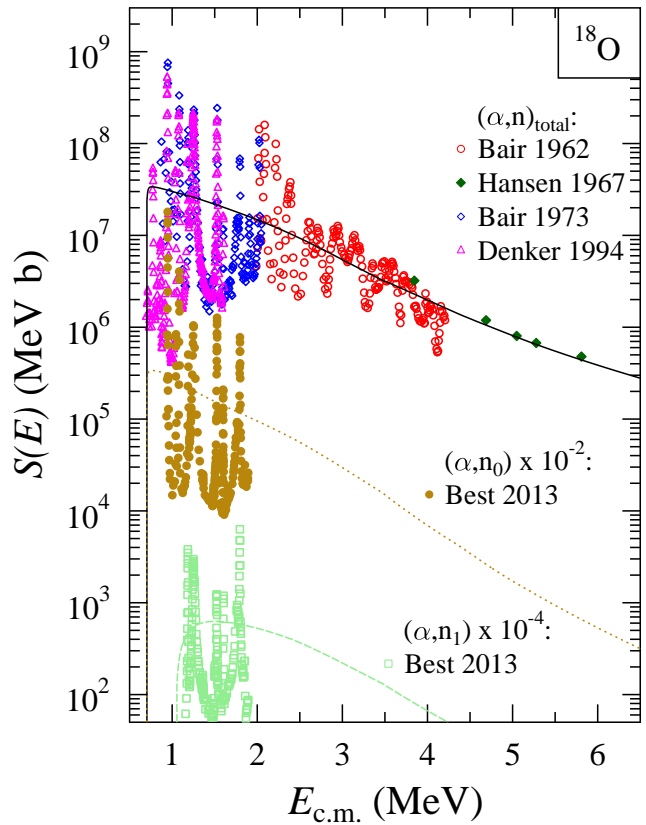}
\caption{
\label{fig:o18an}
(Color online)
Total \oviii \ran \nei\ \sfact\ from neutron counting experiments
\cite{Bair62,Han67,Bair73,Denker94} and partial \rannull\ and \rani\ cross
section measurements \cite{Best13_o18} in comparison to TALYS
calculations. The \rannull\ and \rani\ data are scaled by factors of $10^{-2}$
and $10^{-4}$ for better visibility. Further discussion see text.
}
\end{figure}

The present study starts from the first NACRE compilation \cite{NACRE} in
1999; the updated NACRE-2 compilation \cite{NACRE2} ends at $A = 16$ and does
not include the reactions under analysis. NACRE lists four experiments where
total \oviii \ran \nei\ cross sections were measured by neutron
counting. This technique does not provide much information on the
neutron energy which complicates a precise calibration of the energy-dependent
efficiency of the neutron longcounters. In addition, resonances in background
reactions may be misinterpreted. Nevertheless, the four data sets by Bair and
Willard \cite{Bair62} (hereafter: Bair62; the other data sets are referenced
by the first author in the following), Hansen {\it et al.}\ \cite{Han67}, Bair
and Haas \cite{Bair73}, and Denker \cite{Denker94} are in reasonable
agreement (see Fig.~\ref{fig:o18an}). The data cover energies from close above
the reaction threshold up to about 10 MeV. Note that the \oviii \ran
\nei\ reaction has a slightly negative Q-value of $Q = -698$ keV, leading to a
threshold at $E_\alpha = 842$ keV in the laboratory system.

The TALYS calculations show good agreement with the Hansen data at higher
energies above 5 MeV, and the calculations reproduce the average trend of the
Bair62 data down to about 2 MeV. At even lower energies, there is still
reasonable agreement between the calculation and the average trend of the Bair
data and the Denker data.

In addition to the longcounter data, time-of-flight (TOF) data have been
used by Hansen {\it et al.}\ \cite{Han67} at higher energies around 10 MeV to
discriminate between the final states in the residual \nei\ nucleus. The
resolution was not sufficient to resolve all individual levels of \nei ; only
angular distributions for three groups (n$_0$+n$_1$; n$_2$+n$_3$;
n$_5$+n$_6$+n$_7$) are shown in \cite{Han67}. The angle-integrated cross
sections of these groups are in reasonable agreement with the TALYS
calculations with deviations below a factor of two in all cases. In
particular, the (n$_0$+n$_1$) group with cross sections of about 80 to 45 mb
from 9.8 to 12.3 MeV are nicely reproduced within about 20\%, giving some
confidence in the calculated branching ratios to the lowest states in \nei .

The latest study by Best {\it et al.}\ \cite{Best13_o18} at low energies
improves the previous longcounter measurements by an additional determination
of the \oviii \rani \nei\ cross section by \g -ray spectroscopy. The
information from the \g -ray data on the \rani\ cross section is used to
calculate the contribution of the \rani\ channel to the total neutron yield
which is measured as in the other studies by neutron counting. After
subtraction of the \rani\ yield, the remaining yield is assigned to the
\rannull\ channel (other channels are closed for the low energies under study
in \cite{Best13_o18}), and this remaining yield is then converted to the
\rannull\ cross section with smaller uncertainties because the neutron energy
in the \rannull\ channel is now known from kinematics; thus, the neutron
detection efficiency can be determined with improved accuracy.

The total \oviii \ran \nei\ cross section of the Best data is in good
agreement with the other data sets, and also the branching ratio between the
\rannull\ and \rani\ cross sections is on average well reproduced by the TALYS
calculations (see Fig.~\ref{fig:o18an}); obviously, the branching ratio of
the individual resonances cannot be reproduced by the StM calculations. This
leads to three conclusions for 
the \oviii \ran \nei\ reaction: First, this cross section is well determined
experimentally from several data sets which agree with each other
\cite{Bair62,Han67,Bair73,Denker94,Best13_o18}. Second, the statistical model
is able to predict the average cross section for both open channels at low
energies. Third, the excellent performance of the simple \al -nucleus
potential of McFadden and Satchler \cite{McF66} in the $20 \le A \le 50$ mass
range \cite{Mohr15} can at least be extended down to \oviii .

\section{\ovii ($\alpha$,${\rm{n}}$)\nenull }
\label{sec:o17}
From the above conclusions on the \oviii \ran \nei\ reaction, similar findings
are expected for the \ovii \ran \nenull\ reaction because the same
experimental techniques have been applied by the same groups. However, this is
not the case. The available experimental data are in part contradictory for
the \ovii \ran \nenull\ reaction.

\begin{figure}[htb]
\includegraphics[width=0.99\columnwidth,bbllx=3,bblly=0,bburx=246,bbury=313,clip=]{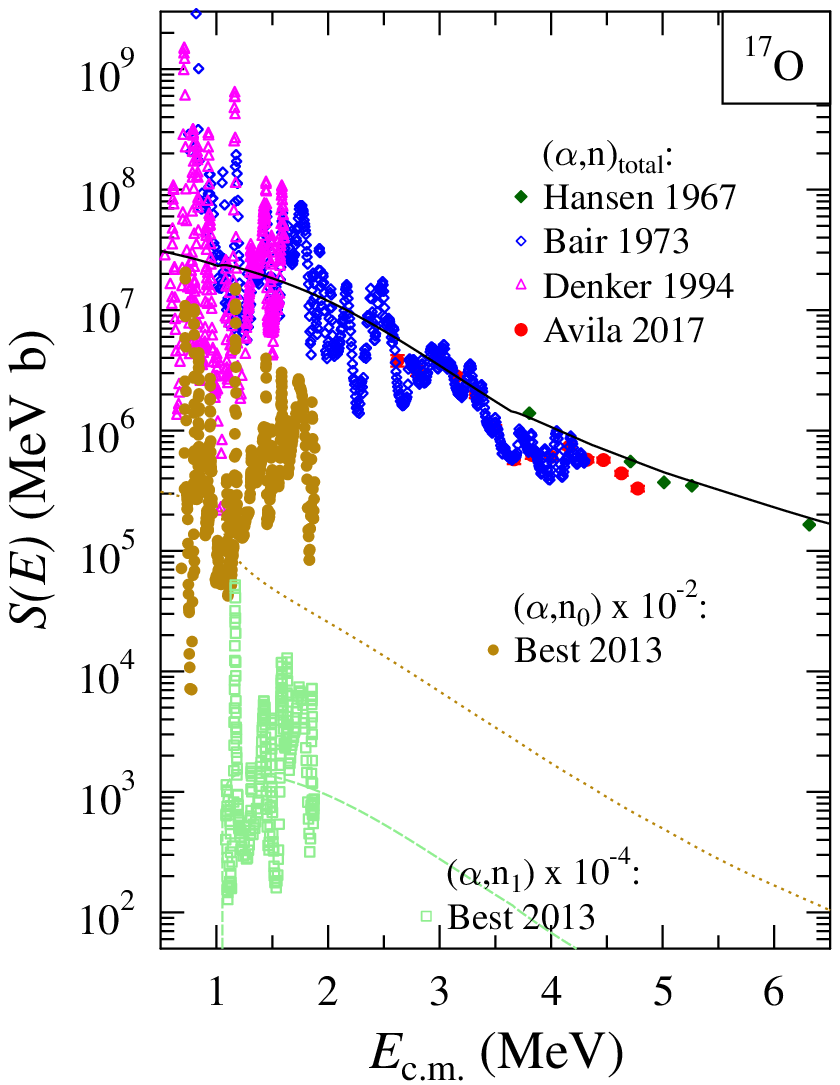}
\caption{
\label{fig:o17an}
(Color online)
Total \ovii \ran \nenull\ \sfact\ from neutron counting experiments
\cite{Han67,Bair73,Denker94} and partial \rannull\ and \rani\ cross
section measurements \cite{Best13_o18} in comparison to TALYS calculations
(total \ran : full black line; \rannull : golden dotted; \rani : lightgreen
dashed). Further total \ran\ data are measured by detection of the
\nenull\ recoil nucleus in inverse kinematics \cite{Avi17}. The \rannull\ and
\rani\ data are scaled by factors of $10^{-2}$ and $10^{-4}$ for better
visibility. Further discussion see text.
}
\end{figure}

Similar to \oviii \ran \nei , the starting point of the present analysis
is the NACRE compilation from 1999. Three data sets are listed, starting with
the early work of Hansen {\it et al.}\ \cite{Han67}, the data by Bair and Haas
\cite{Bair73}, and the unpublished data by Denker \cite{Denker94}. All
experiments use simple neutron counting techniques. Similar to the
\oviii\ case, the three data sets are in reasonable agreement (see
Fig.~\ref{fig:o17an}). Very recently, Avila {\it et al.}\ \cite{Avi17} have
measured the \ovii \ran \nenull\ reaction in inverse kinematics by the
detection of the residual \nenull\ nucleus for energies corresponding to
$E_\alpha \approx  3 - 6$ MeV. Also these data with their completely different
systematic uncertainties agree well in the overlap regions with the Bair data
and the Hansen data. The TALYS calculation is able to reproduce the
experimental data at higher energies. At lower energies the cross section is
dominated by individual resonances, but still the statistical model
calculations reproduce the average trend of the data.

Again similar to the \oviii\ case, Hansen {\it et al.}\ \cite{Han67} have
applied the TOF technique to discriminate between the final states in \nenull
.  Angular distributions for the n$_1$, the n$_2$, and the sum of the
(n$_4$+n$_5$) channels are shown for energies between 9.8 and 12.3 MeV, and it
is pointed out that the n$_0$ channel and the n$_3$ channel are only weakly
populated, thus preventing an analysis. The experimental data points for the
n$_1$ and n$_2$ channels (from 9.8 to 12.3 MeV: about 80 to 30 mb for the
n$_1$ channel and 90 to 70 mb for the n$_2$ channel) are reproduced by the
TALYS calculations with deviations below about 20\%, and the calculated n$_0$
cross section is about a factor of five lower than the n$_1$ channel. This
confirms the TALYS calculations for the branchings to individual final states
in \nenull .

\begin{figure}[htb]
\includegraphics[width=0.99\columnwidth,bbllx=3,bblly=0,bburx=254,bbury=274,clip=]{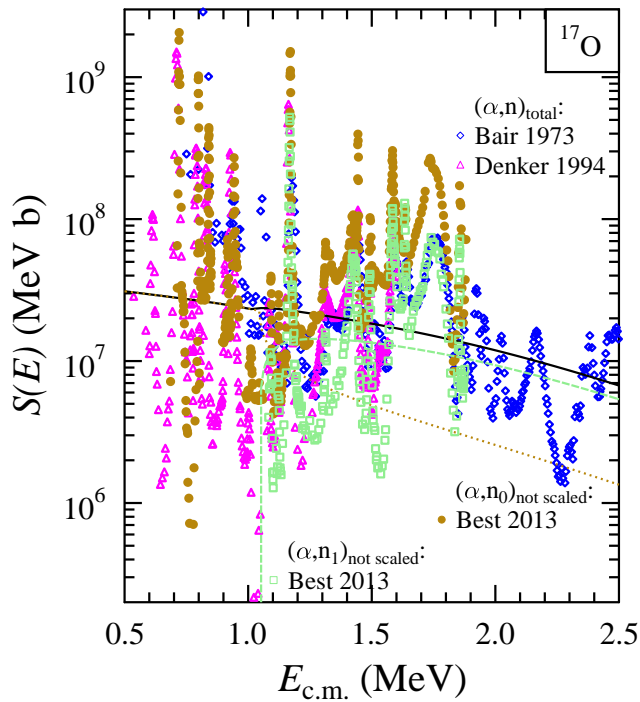}
\caption{
\label{fig:o17anlow}
(Color online)
Same as Fig.~\ref{fig:o17an} for low energies. Above the opening of the n$_1$
channel at $E_\alpha = 1293$ keV ($E_{\rm{c.m.}} = 1047$ keV), the Best data
for the n$_0$ channel exceed the Bair and the Denker data and are also
significantly above the TALYS prediction (golden dotted). Below the n$_1$
threshold, the experimental data sets roughly agree. Note that contrary to
Fig.~\ref{fig:o17an}, the \rannull\ and \rani\ data are not scaled. Further
discussion see text. 
}
\end{figure}

Again similar to the \oviii\ case, Best {\it et al.}\ \cite{Best13_o17} have
extended the neutron counting experiments by an additional measurement of the
\ovii \rani \nenull\ reaction by \g -spectroscopy of the 1634 keV \g -ray from
the decay of the first excited state in \nenull\ to the ground state. The
\rani\ data at low energies are on average well reproduced by the TALYS
calculations. Then, Best {\it et al.}\ calculate the yield of the
\rani\ reaction in their neutron detector, and from the difference of the
measured yield and the calculated \rani\ yield the \rannull\ cross section is
extracted. Other channels are closed at the energies under study in
\cite{Best13_o17}. Contrary to the TOF results by Hansen {\it et
  al.}\ \cite{Han67} around 10 MeV, it is found for the low energy region that
the n$_0$ channel is dominating over the n$_1$ channel. As can be seen from
Fig.~\ref{fig:o17an}, the TALYS calculation is significantly lower than the
experimental result for the \rannull\ channel. Interestingly, this discrepancy
between the Best data for the n$_0$ channel and the TALYS calculation appears
mainly in the energy region above the opening of the n$_1$ channel (see
Fig.~\ref{fig:o17anlow}).

It is somewhat difficult to visualize the essential discrepancies between the
various experimental data sets; an attempt is made in
Fig.~\ref{fig:o17anlow}. Above the n$_1$ threshold at $E_{\rm{c.m.}} = 1047$ keV
and clearly visible above about 1.3 MeV, the Best \rannull\ data exceed the
total \ran\ data of Bair and of Denker by about a factor of three. The TALYS
calculation predicts on average a weak n$_0$ channel and a dominating n$_1$
channel whereas the Best data show the opposite trend over the whole energy
range. Note that the TALYS predictions are verified around 10 MeV according to
the Hansen TOF data.

There is an additional experiment by McDonald {\it et al.}\ \cite{McD76} on
isospin-forbidden particle decays in \nei . An attempt was made in
\cite{McD76} to find weak $T = 3/2$ resonances in the \ovii \rannull
\nenull\ channel by neutron detection in an energy-sensitive NE213
scintillator and in the \ovii \rani
\nenull\ channel by \g -spectroscopy. As a byproduct, neighboring strong
$T=1/2$ resonances were also seen in \cite{McD76}. In particular, two
resonances are discussed explicitly in \cite{McD76}.

At $E_{\rm{c.m.}} = 1491$ keV a resonance was found in the \rani\ channel, but no
enhanced yield was seen in the \rannull\ channel. $\Gamma_{{\rm{n}}_0}/\Gamma
< 0.3$ was deduced from the data, in conflict with the Best data which give
$\Gamma_{{\rm{n}},0} = 5.13$ keV and $\Gamma_{{\rm{n}},1} = 3.05$ keV, leading
to $\Gamma_{{\rm{n}},0}/\Gamma_{{\rm{n}},1} = 1.68$. For completeness it
should be noted that the Denker data for the total \ran\ cross section agree
almost perfectly with the Best data for the \rani\ channel around the 1491
keV resonance.

For the tail of the broad resonance at $E_{\rm{c.m.}} = 1753$ keV ($E_\alpha =
2165$ keV) differential cross sections of
slightly below 2 mb/sr are given at $E_\alpha \approx 2200$ keV for the
\rannull\ and the \rani\ channels in \cite{McD76}. Assuming isotropy, this
corresponds to angle-integrated cross sections of the order of $20-25$
mb. Interestingly, the \rani\ cross section is in rough agreement with the
Best data, but the \rannull\ cross section is again significantly lower (about
a factor of two) than the Best result.

\begin{figure}[htb]
\includegraphics[width=0.99\columnwidth,bbllx=3,bblly=0,bburx=246,bbury=282,clip=]{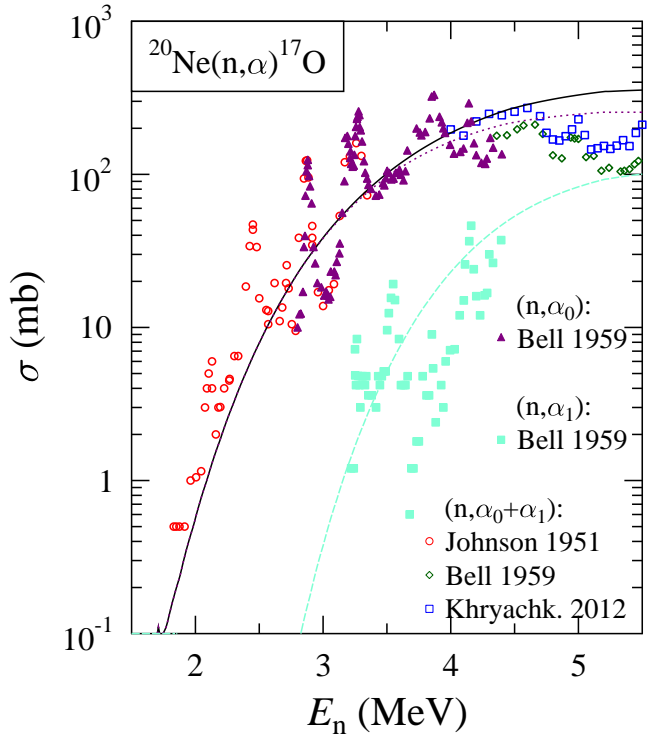}
\caption{
\label{fig:ne20na}
(Color online)
Experimental cross section of the \nenull \rna \ovii\ reaction
\cite{John51,Bell59,Khr12a,Khr12b} in comparison to a statistical model
calculation. Because of the dominating \al $_0$ channel, the \rna\ data
provide an additional constraint for the \ovii \rannull \nenull\ cross
section.
}
\end{figure}

The reverse \nenull \rna \ovii\ reaction can be used to provide a further
constraint on the \ovii \rannull \nenull\ data. This possibility was
unfortunately disregarded in the previous works \cite{NACRE,Best13_o17}. Three
data sets are available for neutrons in the low MeV energy region. The early
data by Johnson {\it et al.}\ \cite{John51} are composed of the \al $_0$ and
\al $_1$ channels and cover the low-energy region. Bell {\it et
  al.}\ \cite{Bell59} are able to resolve the \al $_0$ and \al $_1$ channels
for low neutron energies between 3 and 4.5 MeV; at higher energies also only
the sum of the first two channels is reported. At energies above 4 MeV,
recently Khryachkhov {\it et al.}\ \cite{Khr12a,Khr12b} have also measured the
sum of the \al $_0$ and \al $_1$ channels. The \rna\ data are in good
agreement below 5 MeV (see Fig.~\ref{fig:ne20na}) although at higher energies
discrepancies up to about a factor of two are found. The Bell data clearly
indicate that the \rnanull\ channel is dominating with a minor contribution of
the \rnai\ channel of the order of 10\%. Higher-lying final states in
\ovii\ do not play a role at energies below 5 MeV.

The TALYS calculations are able to reproduce the average trend of the
experimental data. The agreement is very good for the \rnanull\ channel
between 3 and 4 MeV, corresponding to slightly lower energies $E_\alpha$ in
the \ran\ reaction because of the small negative $Q$-value of the
\rna\ reaction of $Q = -587$ keV. 

\begin{figure}[htb]
\includegraphics[width=0.99\columnwidth,bbllx=3,bblly=0,bburx=246,bbury=282,clip=]{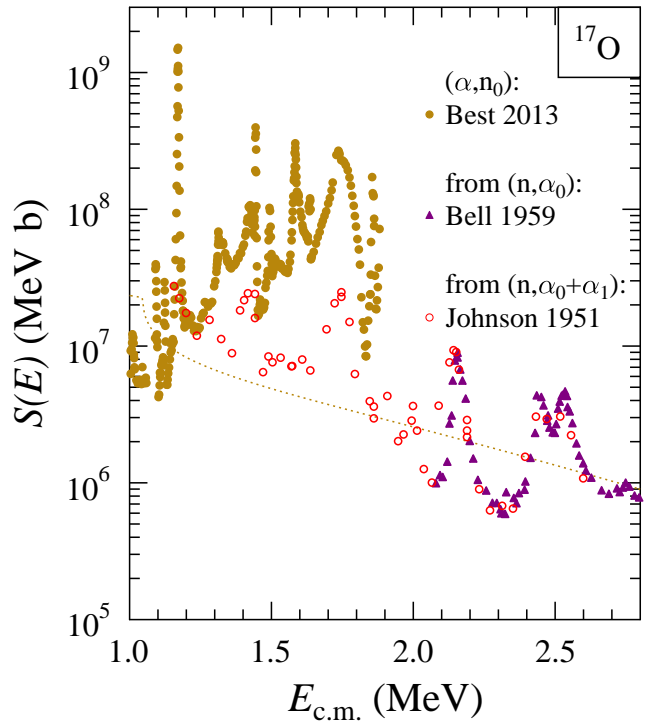}
\caption{
\label{fig:o17ancompare}
(Color online)
Same as Fig.~\ref{fig:o17an} for the overlap region between the Best data for
the \rannull\ channel and the converted \rna\ data of Johnson {\it et
  al.}\ \cite{John51} and Bell {\it et al.}\ \cite{Bell59}. The \rannull\ data
are in significant disagreement to the \rna\ data. The TALYS calculation
reproduces the average trend of the \rna\ data.
Further discussion see text.
}
\end{figure}

The \nenull \rnanull \ovii\ data can be converted to \ovii \rannull
\nenull\ cross sections by application of the reciprocity theorem. The
comparison between the Best data for the \rannull\ channel and the converted
\rna\ data of Johnson and Bell is shown in Fig.~\ref{fig:o17ancompare}. A
significant discrepancy can be seen between the Best data and the data from
the reverse \rna\ reaction. For completeness also the TALYS calculation for
the \rannull\ channel is included in Fig.~\ref{fig:o17ancompare} which clearly
favors the lower data from the \rna\ reaction.

Summarizing the above results, there is a clear experimental contradiction
between the Best data for the \rannull\ channel on the one hand and the
\ran\ data of McDonalds and the \rna\ data of Johnson and of Bell on the other
hand. The TALYS calculation clearly favors the lower \rannull\ data of
McDonalds, Johnson, and Bell. The \rannull\ data of Best are also above the
total \ran\ data of Denker and Bair (at least above the
\rani\ threshold). Thus, the simplest approach for consistency is a reduction
of the \rannull\ data of Best above the \rani\ threshold by a significant
amount. Typically, this reduction should be at least a factor of two to three
(but an energy-independent reduction factor may be inappropriate). Any other
solution would require the modification of several data sets which are roughly
consistent with each other. Fortunately, new experiments for the \ovii \ran
\nenull\ reaction are in preparation \cite{ND17} using improved
energy-sensitive neutron detectors \cite{Bec16}.

\section{\fvii ($\alpha$,${\rm{p}}$)\nenull }
\label{sec:f17}
\begin{figure}[htb]
\includegraphics[width=0.99\columnwidth,bbllx=0,bblly=0,bburx=246,bbury=282,clip=]{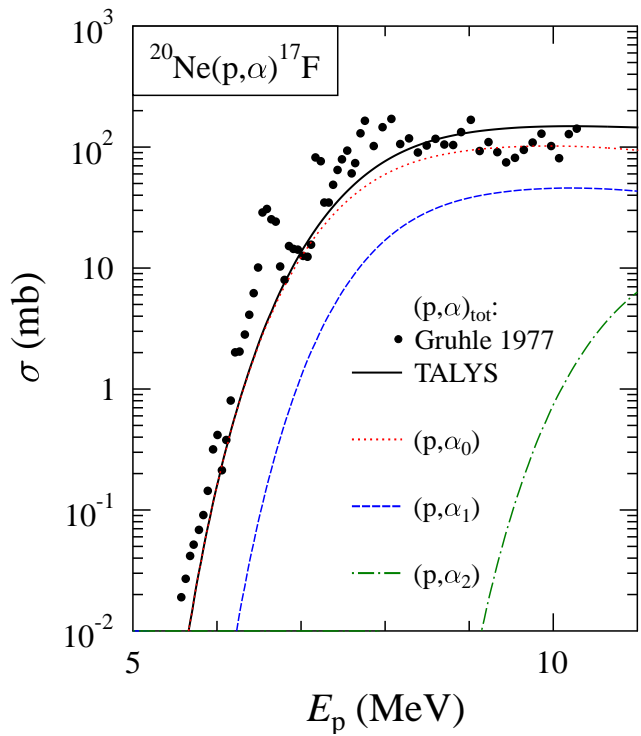}
\caption{
\label{fig:ne20pa}
(Color online)
Experimental cross section of the \nenull \rpa \fvii\ reaction
\cite{Gru77} in comparison to a statistical model calculation. Similar to the
\nenull \rna \ovii\ reaction, because of the dominating \al $_0$ channel the
\rpa\ data provide an additional constraint for the \fvii \rapnull
\nenull\ cross section. 
}
\end{figure}

A detailed discussion of the \fvii \rap \nenull\ reaction will be given in a
forthcoming paper with the upcoming experimental results from ANL
\cite{Avi18}. As a first step, the reverse \nenull \rpa \fvii\ reaction was
studied. Fig.~\ref{fig:ne20pa} shows the experimental results of Gruhle {\it
  et al.}\ \cite{Gru77} for the total \rpa\ cross section. The TALYS
calculation is again able to reproduce the data quite well. 

According to TALYS, the \rpanull\ channel is dominating over the whole energy
range of the Gruhle data with a small contribution ($\lesssim 20\%$) from the
\rpai\ channel and negligible contributions from higher-lying channels like
\rpaii . Such a branching ratio is expected from the negative $Q$-value of the
\rpa\ reaction and the resulting strong Coulomb suppression of the
higher-lying final states in \fvii .

Because of the dominance of the \rpanull\ channel, the experimental \rpa\ data
can be approximately converted to the \fvii \rapnull \nenull\ cross
section. However, a comparison to the upcoming \fvii \rap \nenull\ data is
complicated by the fact that -- according to TALYS -- the \rapnull\ channel is
relatively weak in the \fvii \rap \nenull\ reaction. The predicted branching
ratios and consequences for the analysis of the experimental data will be
presented in \cite{Avi18}. Note that high-lying excited states in
\nenull\ from the \fvii \rap \nenull\ reaction may decay to \ovi\ $+$
\al\ before the residual \nenull\ nucleus can be detected in the MUSIC chamber
at ANL; this complication remained negligible in the analysis of the \ovii
\ran \nenull\ data \cite{Avi17} because of the small $Q$-value of the
\ran\ reaction.

\section{Astrophysical reaction rates}
\label{sec:astro}
The astrophysical reaction rate \Nsv\ is essentially an average cross section
where the averaging is weighted by the thermal Maxwell-Boltzmann distribution
of the colliding nuclei. For a given temperature $T$ (typically given as $T_9
= T / 10^9$ K) the reaction rate \Nsv\ is dominated by the cross section in
the so-called Gamow window which is located around $E_0 = 1150$ keV (1820 keV;
2390 keV) for $T_9 = 1$ ($T_9 = 2$; 3) for \ovii\ and \oviii\ and has a width
$\Delta$ of about 725 keV (1295 keV; 1815 keV) in the center-of-mass system
\cite{Rol88,Ili07}. Note that the simple Gamow window approach does not hold
for the \oviii \ran \nei\ reaction at very low temperatures because of the
negative $Q$-value of about $-0.7$ MeV.

Obviously, the statistical model calculations should be able to provide the
reaction rate \Nsv\ at high temperatures where the cross section in the Gamow
window is composed of a sufficiently high number of resonances. This
definitely holds for energies above about $3-4$ MeV where the averaged cross
sections as e.g.\ measured by Hansen {\it et al.}\ \cite{Han67} or Avila {\it
  et al.}\ \cite{Avi17} show a smooth energy dependence (see
Figs.~\ref{fig:o18an} and \ref{fig:o17an}). Thus, the statistical model is
definitely valid at the corresponding temperatures slightly above $T_9 =
3$. At lower temperatures down to about $T_9 = 1$ still several resonances are
located in the Gamow window. Here \Nsv\ from the statistical model should
remain more or less reliable (say within a factor of two or so) because
\Nsv\ approximately averages over the relatively broad Gamow window. This
reliability of \Nsv\ from the statistical model is confirmed by the reasonable
agreement with the experimental rates which are calculated from the sum over
the contributing resonances for the nuclei under study. However, below $T_9
\approx 1$, \Nsv\ is dominated by very few individual resonances. Here the
statistical model is not able to predict \Nsv\ with sufficient accuracy.

The reaction rate \Nsv\ increases dramatically with temperature. For better
visibility, in the following Figs.~\ref{fig:o18an_rate} and
\ref{fig:o17an_rate} the reaction rates \Nsv\ are normalized to a reference
rate which is taken from the \Nsv\ fit functions of the NACRE compilation for
\ovii\ and \oviii .
The rates will be discussed with a focus on the low temperature region which
corresponds to the astrophysical \spro .

\subsection{\oviii \ran \nei }
\label{sec:o18an_rate}
\begin{figure}[htb]
\includegraphics[width=0.99\columnwidth,clip=]{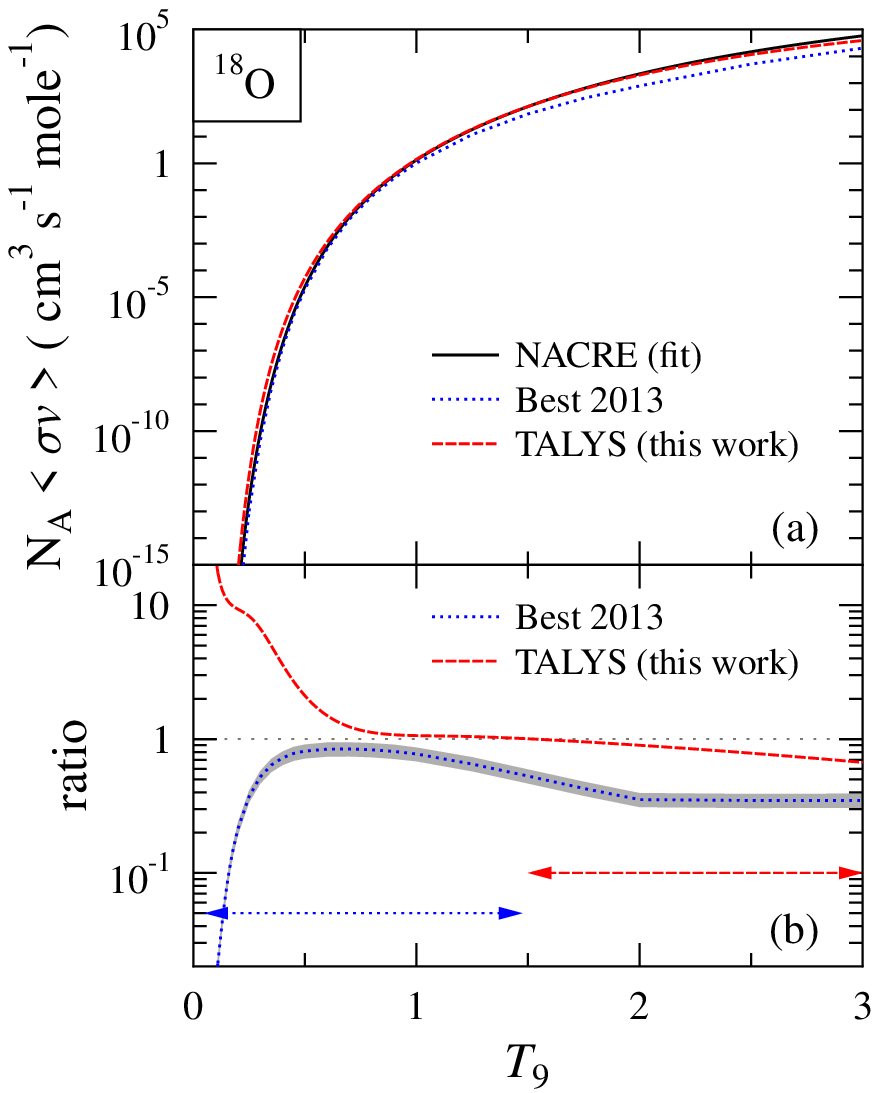}
\caption{
\label{fig:o18an_rate}
(Color online) Reaction rate \Nsv\ of the \oviii \ran \nei\ reaction from the
NACRE compilation, from the experimental resonance properties by Best {\it et
  al.}\ \cite{Best13_o18}, and from the statistical model calculations using
TALYS. The upper part (a) shows the rates \Nsv ; the lower part (b) shows the
rates normalized to the NACRE fit, including horizontal arrows which indicate
the approximate validity of the different rates. Further discussion see text.
}
\end{figure}

The latest calculation of \Nsv\ for the reaction \oviii \ran \nei\ by Best
{\it et al.}\ \cite{Best13_o18} is based on an R-Matrix analysis of the
experimental data which cover an energy range from the threshold up to about 2
MeV; i.e., the Gamow window is fully covered (including the width $\Delta$) up
to $T_9 \approx 1.5$. Therefore, at higher temperatures the rate \Nsv\ is
calculated from the statistical model which has been scaled to \Nsv\ from
experiment at $T_9 = 2$. The Best results are slightly lower than NACRE, but
the deviation does not exceed a factor of two for $0.5 \lesssim T_9 \lesssim
2.0$. The significantly lower \Nsv\ at very low temperatures below $T_9 = 0.5$
results from the fact that the lowest resonance in the Denker data at 888 keV
is considered as spurious and has been assigned to the \ovii \ran
\nenull\ reaction by Best {\it et al.}\ \cite{Best13_o17}. The results are
shown in Fig.~\ref{fig:o18an_rate}.

The TALYS calculation is between the NACRE rate and the Best rate for
temperatures above $T_9 \approx 1$, and it remains closer to the NACRE rate.
As pointed out above, at temperatures above $T_9 \approx 3$ the statistical
model should be fully valid. Two potential explanations (or a combination of
both) can be given for the deviation between the Best rate and the TALYS rate:
($i$) The number of resonances in the \oviii \ran \nei\ reaction may be
accidentially low in the Gamow window for $T_9 = 2$ around 2 MeV, leading to a
scaling factor significantly below 1.0 in \cite{Best13_o18} for the adjustment
of the statistical model calculations. (Unfortunately, this factor is not
provided in \cite{Best13_o18}.) ($ii$) The Gamow window at $T_9 = 2$ is not
fully covered by the experimental cross sections, leading to a slight
underestimation of \Nsv\ at $T_9 = 2$ because of missing contributions from
energies above 2 MeV.

At very low temperatures below $T_9 \approx 0.7$ the limitations of the
statistical model become clearly visible. The statistical model gives an
almost constant astrophysical \sfact\ of $S(E) \approx 3 \times 10^7$ MeV\,b
down to the threshold of the \ran\ reaction. Such a constant \sfact\ leads to
a significantly enhanced reaction rate \Nsv\ which is excluded by the
experimental data of Denker and Best. Note that the negative $Q$-value and the
resulting threshold for the \ran\ reaction lead to a relatively
well-constrained reaction rate \Nsv\ because resonances at very low energies
with their typically tiny (and often not measureable) resonance strengths
cannot exist.

Summarizing the above, the reaction rate \Nsv\ of the \oviii \ran
\nei\ reaction is well-defined down to low temperatures from the Best
data. Except the spurious resonance at 888 keV, the Denker data and also the
Bair and the Hansen data at higher energies show good agreement and thus
confirm the recommended rate by Best {\it et al.}\ \cite{Best13_o18}. The TALYS
calculation cannot be used below $T_9 \approx 0.7$. Above $T_9 \approx 2$ the
TALYS calculation gives slightly higher \Nsv . Here the TALYS calculation
reproduces the experimental \ran\ data of Bair62, Bair, and Hansen, and it is
close to the evaluation in the NACRE compilation; thus, the TALYS calculation
should be reliable.

\subsection{\ovii \ran \nenull }
\label{sec:o17an_rate}
Similar to the \oviii \ran \nei\ reaction, Best {\it at al.}\ \cite{Best13_o17}
provide the reaction rate \Nsv\ of the \ovii \ran \nenull\ reaction from a
R-matrix fit to their experimental data in the energy range from 0.7 to 1.9
MeV. Two versions of \Nsv\ are listed in \cite{Best13_o17}. The so-called
experimental rate \Nsv $_{\rm{exp}}$ is calculated from the experimental
resonance strengths (excluding contributions from resonances outside the
experimental energy range from 0.7 to 1.9 MeV). The recommended rate \Nsv
$_{\rm{rec}}$ additionally includes estimates for low-lying resonances, and at
higher temperatures the result of a statistical model calculation is
recommended which has been adjusted to the experimental \Nsv $_{\rm{exp}}$ at
$T_9 = 2$ (as in the case of the \oviii \ran \nei\ reaction).

\begin{figure}[htb]
\includegraphics[width=0.99\columnwidth,clip=]{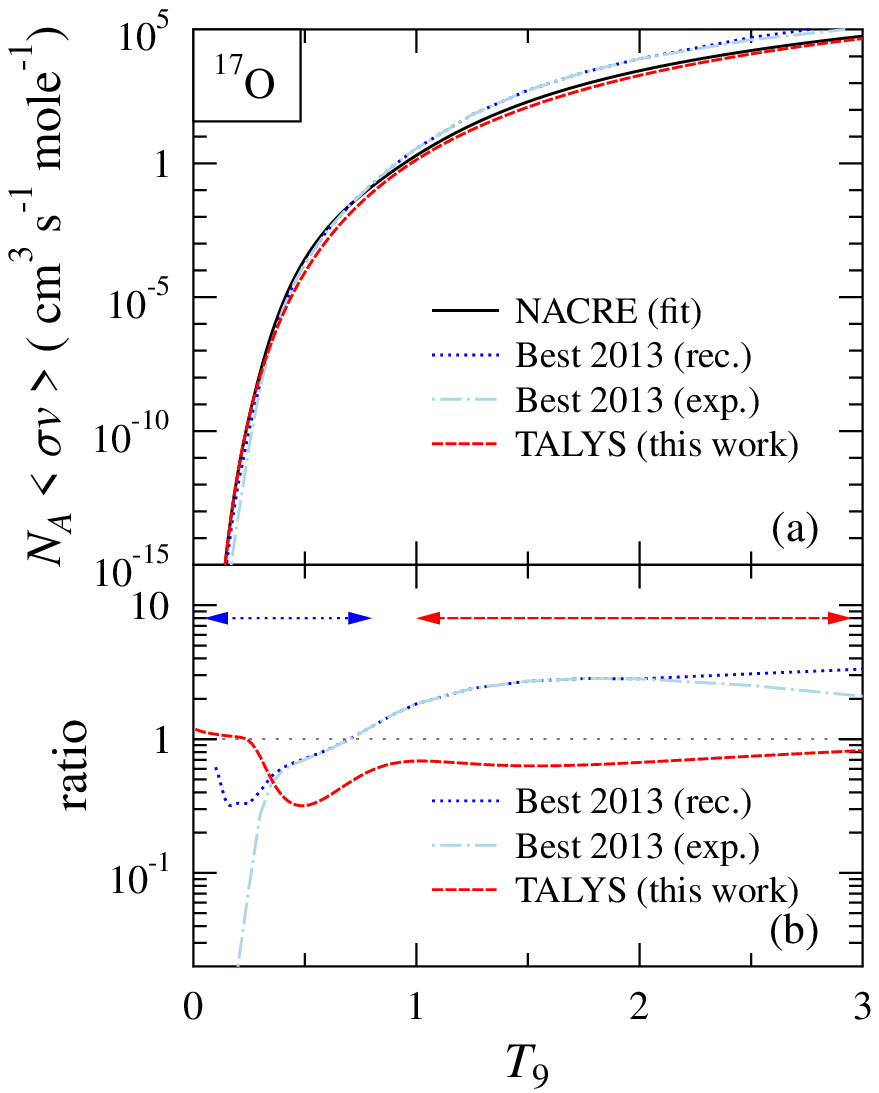}
\caption{
\label{fig:o17an_rate}
(Color online) Reaction rate \Nsv\ of the \ovii \ran \nenull\ reaction from the
NACRE compilation, from the experimental resonance properties by Best {\it et
  al.}\ \cite{Best13_o17}, and from the statistical model calculations using
TALYS. The upper part (a) shows the rates \Nsv ; the lower part (b) shows the
rates normalized to the NACRE fit, including horizontal arrows which indicate
the approximate validity of the different rates. Further discussion see text.
}
\end{figure}

Fig.~\ref{fig:o17an_rate} shows the results. As discussed in
Sec.~\ref{sec:o17}, the Best cross sections are significantly above the other
data from literature in particular for the \rannull\ channel above the
\rani\ threshold. This leads to an enhanced \Nsv\ by about a factor of three
for $1 \le T_9 \le 2$. As expected, above $T_9 = 2$ the enhancement of \Nsv
$_{\rm{exp}}$ decreases because of missing contributions from outside the
experimental energy range, whereas \Nsv $_{\rm{rec}}$ remains a factor of
three above the NACRE rate. The TALYS calculation agrees almost perfectly with
the NACRE compilation, and the temperature dependence is almost identical to
\Nsv $_{\rm{rec}}$ of Best {\it et al.}\ \cite{Best13_o17}, at least for
temperatures above $T_9 = 1$.

At very low temperatures, \Nsv\ is governed by few low-lying resonances which
have not been seen in \ran\ experiments to date. NACRE extends the lowest
experimental \sfact\ data using a constant $S(E)$ down to $E =
0$. Consequently, the experimental \Nsv $_{\rm{exp}}$ by Best {\it et al.}\ is
by far below the NACRE result and the TALYS calculation. The recommended \Nsv
$_{\rm{rec}}$ is closer to the NACRE rate, but still about a factor of two
lower. The agreement between the TALYS calculation and NACRE for the lowest
temperatures is not surprising because the calculated TALYS \sfact\ towards $E
\approx 0$ is close to the chosen constant \sfact\ of NACRE. The rough
agreement between the Best recommendation and the TALYS rate at low
temperatures must be considered as somewhat accidential. However, as the Best
recommended \Nsv $_{\rm{rec}}$ is based on well-chosen average properties of
several unobserved low-lying resonances, the resulting \Nsv $_{\rm{rec}}$
should not deviate by orders of magnitude from a statistical model calculation
which is also based on average properties.  For completeness it can be noted
that a microscopic calculation of the \ovii \ran \nenull\ cross section at low
energies \cite{Des93} gives a rate \Nsv\ which is more than one order of
magnitude lower than the NACRE recommendation at $T_9 \approx 0.1$
\cite{Denker94}.

In summary, the reaction rate \Nsv\ of the \ovii \ran \nenull\ reaction has
significant uncertainties. At low temperatures ($T_9 \lesssim 0.7$) the
recommended rate \Nsv $_{\rm{rec}}$ of Best {\it et al.}\ is a good
choice. Here improved experimental resonance strengths for the yet unobserved 
low-lying resonances could reduce the uncertainties. However, above $T_9 = 1$
the contradictory experimental data lead to uncertainties of at least a factor
of three. Here the Best recommended \Nsv $_{\rm{rec}}$ should be considered as an
upper limit for \Nsv , and the lower limit for \Nsv\ should be taken from the
NACRE compilation or from the present TALYS calculation. A reduction of this
uncertainty requires the resolution of the contradictory experiments.

\section{Sensitivity to the chosen \al -nucleus potential}
\label{sec:alpha}
It has been shown that the calculation of cross sections of \ran\ cross
sections in the $20 \le A \le 50$ mass range is mainly sensitive to the
chosen \al -nucleus potential \cite{Mohr15}. This also holds for the present
study for \ovii\ and \oviii\ where the \rap\ channel remains closed up to more
than 5 MeV, thus minimizing the role of the nucleon-nucleus potential. Whereas
in the in the $20 \le A \le 50$ mass range the McFadden/Satchler potential
\cite{McF66} provides good results \cite{Mohr15} and a relative small
sensitivity of the reaction cross sections on the \al -nucleus potential was
seen recently for $^{64}$Zn \cite{Orn16}, for heavy targets typically huge
deviations are found from calculations of different \al -nucleus potentials
(e.g., \cite{Som98,Gal05,Sau11}).

\begin{figure}[htb]
\includegraphics[width=0.99\columnwidth,clip=]{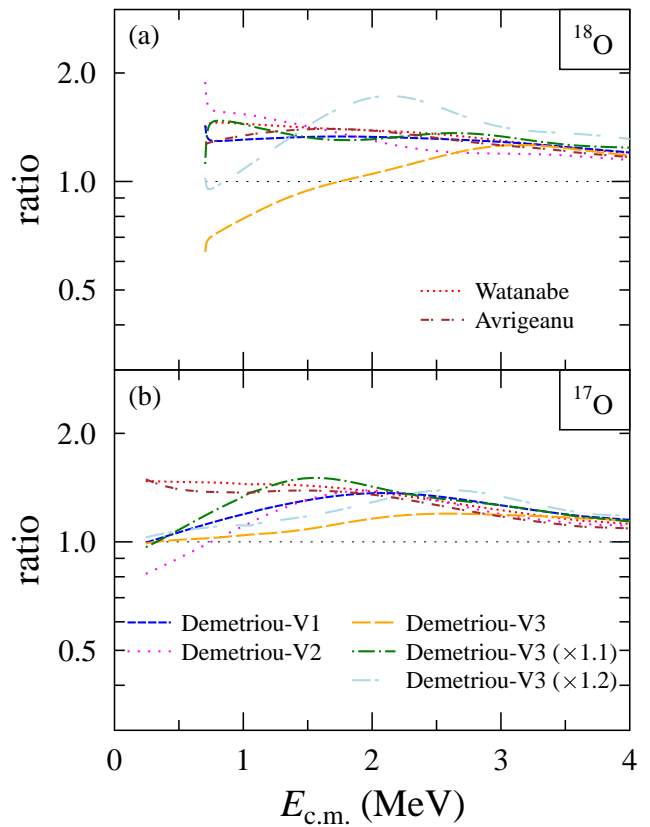}
\caption{
\label{fig:aomp_ratio}
(Color online)
Ratio between the calculated \ran\ cross sections, normalized to the standard
potential of McFadden and Satchler \cite{McF66}, using different \al -nucleus
potentials for \oviii\ (a) and \ovii\ (b). The calculated cross
sections do not vary by more than a factor of two.
}
\end{figure}

Excitation functions for the \oviii \ran \nei\ and \ovii \ran
\nenull\ reactions were calculated from several \al -nucleus potentials. For
presentation, the calculated excitation functions are normalized to the
reference calculation using the McFadden/Satchler potential \cite{McF66} (see
Fig.~\ref{fig:aomp_ratio}).

The following potentials were investigated. The TALYS default potential is
based on Watanabe \cite{Wat58} and results in slightly higher cross
sections. Similar findings are obtained from Avrigeanu {\it et
  al.}\ \cite{Avr14} which will be used as default in the next TALYS
versions. Three different versions, provided by Demetriou {\it et al.}
\cite{Dem02}, are also shown in Fig.~\ref{fig:aomp_ratio}. Whereas the first
two versions give cross sections slightly above McFadden/Satchler, the third
version is slightly lower in particular for \oviii\ at low energies.

Recently, it has been suggested to multiply the real potential of the third
version of Demetriou {\it et al.}\ by a factor of $1.1 - 1.2$
\cite{Net15}. Later, the same factor has been applied in
\cite{Sch16}, and good agreement for several reactions was found (see
Supplement of \cite{Sch16}). The corresponding calculations using the
Demetriou-V3 potential multiplied by 1.1 or 1.2 are also slightly
higher than the McFadden/Satchler result.

For completeness, it has to be noted that most of the global potentials
\cite{Dem02,Mohr13,Avr14,Sch16} have been optimized for medium-mass and heavy
targets which may lead to additional uncertainties for the light targets under
study in this work. For the ATOMKI-V1 potential \cite{Mohr13} it is explicitly
stated that it is applicable only above $A \gtrsim 60$; thus, the results from
ATOMKI-V1 are not included in Fig.~\ref{fig:aomp_ratio}.

Usually, at higher energies different \al -nucleus potentials show a
trend to provide almost identical reaction cross sections with small
deviations of the order of $10-20$\%. Such a convergence is already found at
about 4 MeV for \oviii \ran \nei\ and \ovii \ran \nenull\ (see
Fig.~\ref{fig:aomp_ratio}). But interestingly also at lower energies the
differences from the various \al -nucleus potentials remain quite limited
within about a factor of two.

\section{Conclusions}
\label{sec:conc}
The present study shows that statistical model calculations in combination
with the simple \al -nucleus potential by McFadden/Satchler are able to
reproduce the cross sections of \al -induced reactions even for light nuclei
with masses $A \lesssim 20$. Obviously, this result holds mainly for higher
energies above a few MeV. At lower energies the statistical model cannot
reproduce the individual resonances, but is still able to reproduce the
average trend of the energy dependence which is essential for the prediction
of astrophysical reaction rates \Nsv . These results extend the conclusions of 
\cite{Mohr15} towards lighter nuclei. The results from other recent \al
-nucleus potentials do not differ by more than a factor of two from the widely
used McFadden/Satchler potential.

The statistical model calculations can be used to predict astrophysical
reaction rates \Nsv\ for higher temperatues above $T_9 \approx 2-3$ with high
reliablity. However, as expected, at very low temperatures below $T_9 \approx
1$ the statistical model predictions are not reliable because the reaction
rates are governed here by the properties of very few individual resonances.

For the \oviii \ran \nei\ reaction good agreement between all experimental
data is found, leading to an experimentally well-constrained reaction rate
\Nsv . Surprisingly, for the \ovii \ran \nenull\ reaction a significant
discrepancy has been found at energies above the \rani\ threshold between the
data by Best {\it et al.}\ \cite{Best13_o17} and several other
\ran\ \cite{Han67,Bair73,McD76,Denker94,Avi17} and
\rna\ \cite{John51,Bell59,Khr12a,Khr12b} data sets. This experimental
discrepancy has to be resolved for a better definition of the \ovii \ran
\nenull\ rate at higher temperatures above $T_9 \approx 1$. For the
astrophysically most important low temperatures below $T_9 = 1$ which are
typical for the \spro , the recommendations of Best {\it et
  al.}\ \cite{Best13_o17} remain valid.

\acknowledgments
I thank M.\ Avila and K.\ E.\ Rehm for motivating this study and for providing
their preliminary data for \fvii . Encouraging discussions with A.\ Best,
R.\ deBoer, and M.\ Wiescher are gratefully acknowledged.
This work was supported by OTKA (K108459 and K120666).

\end{document}